\documentclass[prc,twocolumn,superscriptaddress,showpacs,preprintnumbers,amsmath,amssymb,floatfix]{revtex4}

\usepackage{graphicx}
\usepackage{dcolumn}
\usepackage{bm}
\usepackage{longtable}

\begin{document}

\title{Unfolding the Effects of the T=0 and T=1 Parts of the Two-Body 
Interaction on Nuclear Collectivity in the f-p Shell}
%\title{Insensitivity of the Yrast Spectra of Even-Even Nuclei to the
%T=0 two-body interaction matrix elements}

\author{Shadow J.Q. Robinson}

\affiliation{Department of Physics,
University of Southern Indiana,
Evansville, Indiana 47712}
\author{Alberto Escuderos}
\author{Larry Zamick}
\affiliation{Department of Physics and Astronomy,
Rutgers University, Piscataway, 
New Jersey  08855}

\date{\today}

\begin{abstract}
Calculations of the spectra of various even--even nuclei in the fp
shell ($^{44}$Ti, $^{46}$Ti, $^{48}$Ti, $^{48}$Cr and $^{50}$Cr) are
performed with two sets of two-body interaction matrix elements. The
first set consists of the matrix elements of the FPD6 interaction. The
second set has the same T=1 two-body matrix elements as the FPD6
interaction, but all the T=0 two-body matrix elements are set equal to
zero (T0FPD6). Surprisingly, the T0FPD6 interaction gives a semi-reasonable
spectrum (or else this method would make no sense). A consistent feature for
even--even nuclei, e.g. $^{44,46,48}$Ti and $^{48,50}$Cr, is that the
reintroduction of T=0 matrix elements makes the spectrum look more rotational
than when the T=0 matrix elements are set equal to zero. A common 
characteristic of the results is that, for high spin states, the excitation 
energies are too high for the full FPD6 interaction and too low for T0FPD6,
as compared with experiment. The odd--even nucleus $^{43}$Ti and the odd--odd
nucleus $^{46}$V are also discussed. For $^{43}$Sc the T=0 matrix elements are
responsible for staggering of the high spin states. In general, but not always,
the inclusion of T=0 two-body matrix elements enhances the B(E2) rates.

%Despite the drastic differences between the two interactions,
%the spectra they yield are, while by no means identical, surprisingly
%similar. That the results for the yrast spectra are insensitive to the
%presence or absence of T=0 two-body matrix elements is surprising
%because the only bound two nucleon system has T=0, namely the
%deuteron.  Electric quadrupole transition rates are also examined. It
%is found that the reintroduction of T=0 matrix elements leads to an
%enhancement of B(E2)'s for lower spin transitions. As a counterpoint
%we also discuss the odd A nucleus $^{43}$Sc and the odd-odd nucleus
%$^{46}$V.
\end{abstract}

\maketitle

\section{Introduction}

The study of neutron--proton pairing, especially in the T=0 channel, is a
particularly prominent topic these days. While the number of journal
articles are far too numerous to reference, one might begin to make
some headway into the varied approaches by starting from the
references found in Refs.~\cite{goodmannew,macc1,8488}. In so doing
one will find a field of study filled with disagreement and occasionally
strife.

For example, Macchiavelli et~al.~\cite{macc1} claim that some apparent
indicators of T=0 pairing can really be explained in terms of symmetry
energies. In their abstract they say ``After correcting for the energy
we find that the lowest T=1 state in odd--odd N=Z nuclei is as bound as
the ground state in the neighboring even--even nucleus, thus providing
evidence for isovector np pairing. However the T=0 states in odd--odd
N=Z nuclei are several MeV less bound than the even--even ground
states...there is no evidence for an isoscalar (deuteron like) pair
condensate in N=Z nuclei.''  While in this work we do not want to get
into the arguments between this group and others on this point, we
find their work a useful source of motivation for this present study.

In this work we will examine the yrast spectra of the even--even (fp)
shell nuclei $^{44}$Ti, $^{46}$Ti, $^{48}$Ti, $^{48}$Cr and $^{50}$Cr,
as well as the odd--odd nucleus $^{46}$V. We will perform full fp shell
calculations and compare the spectra to experiment.  For comparison
purposes we also discuss the odd A nucleus $^{43}$Sc ($^{43}$Ti) and
the odd--odd nucleus $^{46}$V. 

In this work we perform the shell model calculations using the shell
model code ANTOINE~\cite{antoine}. In order to best see the effects
of the T=1 and T=0 interactions, we perform two sets of
calculations. In the first we use the FPD6
interaction~\cite{wrichter1}.  Then we do the same calculations but we
set all the T=0 two-body interaction matrix elements to zero. We
shall denote this interaction as T0FPD6. (This modification of an
effective interaction is along the same lines of that used by Satula
et~al. to examine Wigner energies a few years ago~\cite{satula1}.)  We
have used this modification of FPD6 in the past to study a variety of
things and in particular the full fp spectrum of
$^{44}$Ti~\cite{dow1,dow2,dow3,dow4}.

It should be noted that in Refs.~\cite{dow1,dow2,dow3,dow4} a wide
range of topics is addressed beyond the spectra of even--even
nuclei. These topics include a partial dynamical symmetry that arises
when one uses the T0FPD6 interaction in a single j shell for $^{43}$Sc
and $^{44}$Ti. Also, while using the T0FPD6 interaction, a subtle
relationship between the T=$\frac{1}{2}$ states in $^{43}$Sc and the
T=$\frac{3}{2}$ states in $^{43}$Ca can be observed, likewise between the T=$0$
states in $^{44}$Ti and T=$2$ states in $^{44}$Ca.  We also considered
even--odd nuclei and addressed the topic of how the T=0 two-body matrix
elements affect B(M1) transitions---both spin and orbital components,
and Gamow--Teller transitions. In many cases the transition rates were
very sensitive to the presence or absence of the T=0 matrix
elements. This was especially the case for some orbital B(M1)'s and
the Gamow--Teller transitions.

Here things will be kept simple and we focus on the spectra and
B(E2)'s of the yrast levels of selected even--even nuclei.  We will
examine the sensitivity of these observables on the T=0 two-body
interaction matrix elements by setting them to zero and comparing the
results thus obtained with those when the T=0 matrix elements are
reintroduced.

The T0FPD6 interaction is not expected to give good binding energies---clearly 
the T=0 two-body matrix elements are important here. Nor is it
expected to give the relative energies of states of different isospins
in a nucleus. This can be partially compensated by adding a two-body
monopole interaction in the T=0 channel $a+bt(i) \cdot t(j)$, which
for T=0 would be $a(1/4-t(1)\cdot t(2))$. Such monopole interactions have been
studied in the past~\cite{bf64, z65}. However this interaction will not
affect the energy differences of states with the same isospin and it
will not affect the B(E2) rates.

% SHOULD DELETE THIS!!! BUT WHAT ABOUT THE CHASMAN REF?
It should be emphasized that T=0 two-body matrix elements are very
important for binding energies. This is especially made clear by the
schematic models of Chasman where it is shown that both T=0 and T=1
matrix elements are important in describing the Wigner
energy~\cite{chas}. In this work, however, we are focusing on spectra.

\section{Discussion of Some Previous Calculations}

Our entry to this problem considered here was to note that in a single
j shell calculation of $^{44}$Ti the results for the even $J$ states were
almost the same when the T=0 two-body matrix elements of the FPD6
interaction were set equal to zero as they were in a full
calculation. This figure is shown in reference~\cite{dow1}. There is
an offset of the odd $J$ states. However we point out that none of the
odd $J$ states have been found experimentally. In this report we provide
an important source of motivation for experiments that
should be done.

It should be pointed out that in a single j shell calculation (but not
when more than one shell is included) setting the two-body T=0 matrix
elements to a constant will give the same relative spectra of T=0
states in $^{44}$Ti as will be obtained by setting these to zero.

In another vein we showed that when T=0 two-body matrix elements are
set equal to zero one gets a partial dynamical symmetry.  For $I=0$
states of $^{44}$Ti with the following angular momenta $I=3, 7, 9, 10$, and
12, the states can be classified by the dual quantum numbers $(J_p,
J_n)$. However for states with $I=0, 2, 4, 5, 6$, and 8 no such symmetry
exists. We were able to explain this in part by noting that this
symmetry exists only for states with angular momenta which are not
present for a system of identical particles, i.e. $^{44}$Ca.

But even with a full interaction, i.e. when the T=0 matrix elements are
present, the T=0 interaction appears to be weak for the states
$I=3,\ldots, 12\,$ for which the dynamical symmetry exists. For example, the
wave function of the $J=3^+$, T=0 state in an MBZ calculation is
\begin{equation}
\Psi = \sum_{J_P, J_N} D^{I}(J_P J_N) [(j^2)^{J_P}(j^2)^{J_N}]^I
\end{equation}

\begin{center}
\begin{tabular}{llcrcr}
\vspace{0.1cm}
   $J_P$  &  $J_N$ & & $3^+_1$ T=0 & &    $3^+_2$ T=0  \\ \hline
              2    &   2    & &   0.0000      & &  0.0000    \\
              2    &   4    & &   0.6968      & & $-0.1202$  \\
              4    &   2    & &   $-0.6968$   & & 0.1202  \\
              4    &   4    & &   0.0000      & & 0.0000     \\
              4    &   6    & &   0.1202      & & 0.6968     \\
              6    &   4    & &   $-0.1202$   & & $-0.6968$  \\
              6    &   6    & &   0.0000      & & 0.0000     \\
\end{tabular}
\end{center}

The point is that, even with the T=0 interaction present, $(J_p, J_n)$
are almost good quantum numbers. The 3$^+_1$ state consists mostly of
the $(J_p, J_n)$ of (24) and (42); the (46) and (64) amplitudes are only
0.1202. The 3$^+_2$ state is mainly (46) and (64). When the T=0 matrix
elements are turned off, the wave functions collapse to 
\begin{eqnarray}
3^+_1 = \frac{1}{\sqrt{2}} [ (2,4)+(4,2)] \\
3^+_2 = \frac{1}{\sqrt{2}} [ (6,4)+(4,6)]
\end{eqnarray}

Of course intrinsically the T=0 interaction is not weak but it appears
to act weak in certain cases.

\section{Results}

\subsection{The even--even isotopes $^{44,46,48}$Ti and $^{48,50}$Cr 
even $J$ states}

\begin{figure}
\includegraphics[width=85mm,height=4in]{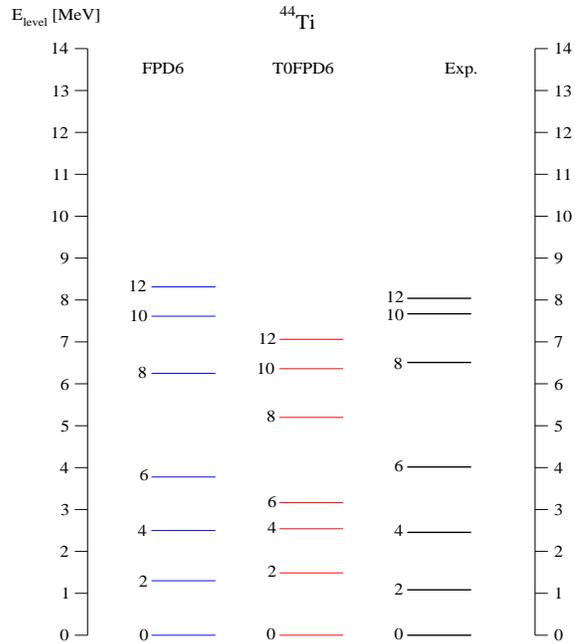}
\caption{ Full fp space calculations of even $J$ T=0 states in $^{44}$Ti and
comparison with experiment.
\label{fig:ti44}
}
\end{figure}

\begin{figure}
\includegraphics[width=85mm,height=4.2in]{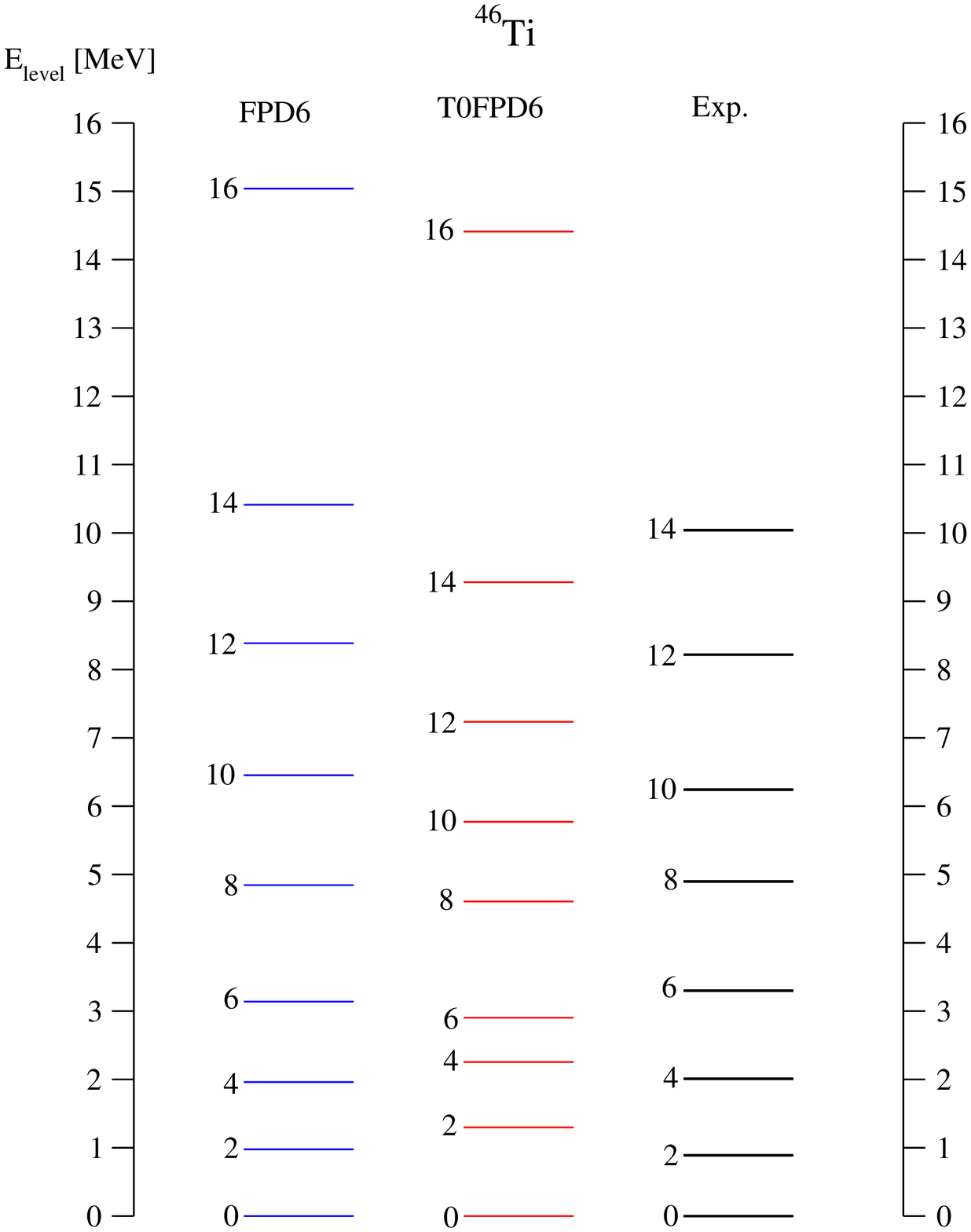}
\caption{Full fp space calculations of even $J$ T=1 states in $^{46}$Ti and
comparison with experiment.
\label{fig:ti46}}
\end{figure}

\begin{figure*}
\includegraphics[angle=270,width=170mm]{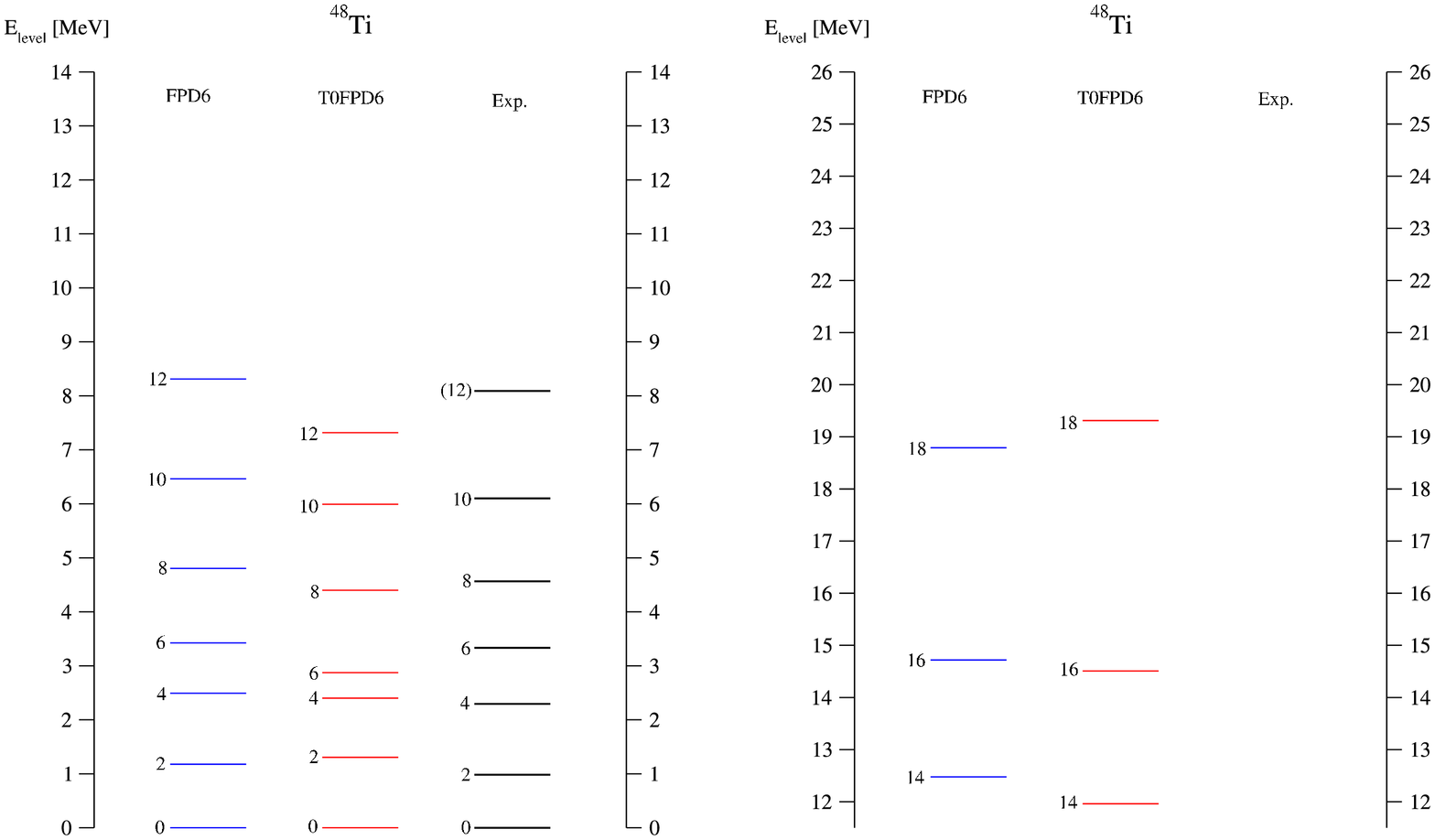}
\caption{Full fp calculations of even $J$ T=2 states in $^{48}$Ti and
comparison with experiment.
\label{fig:ti48}}
\end{figure*}

\begin{figure*}
\includegraphics[angle=270,width=170mm]{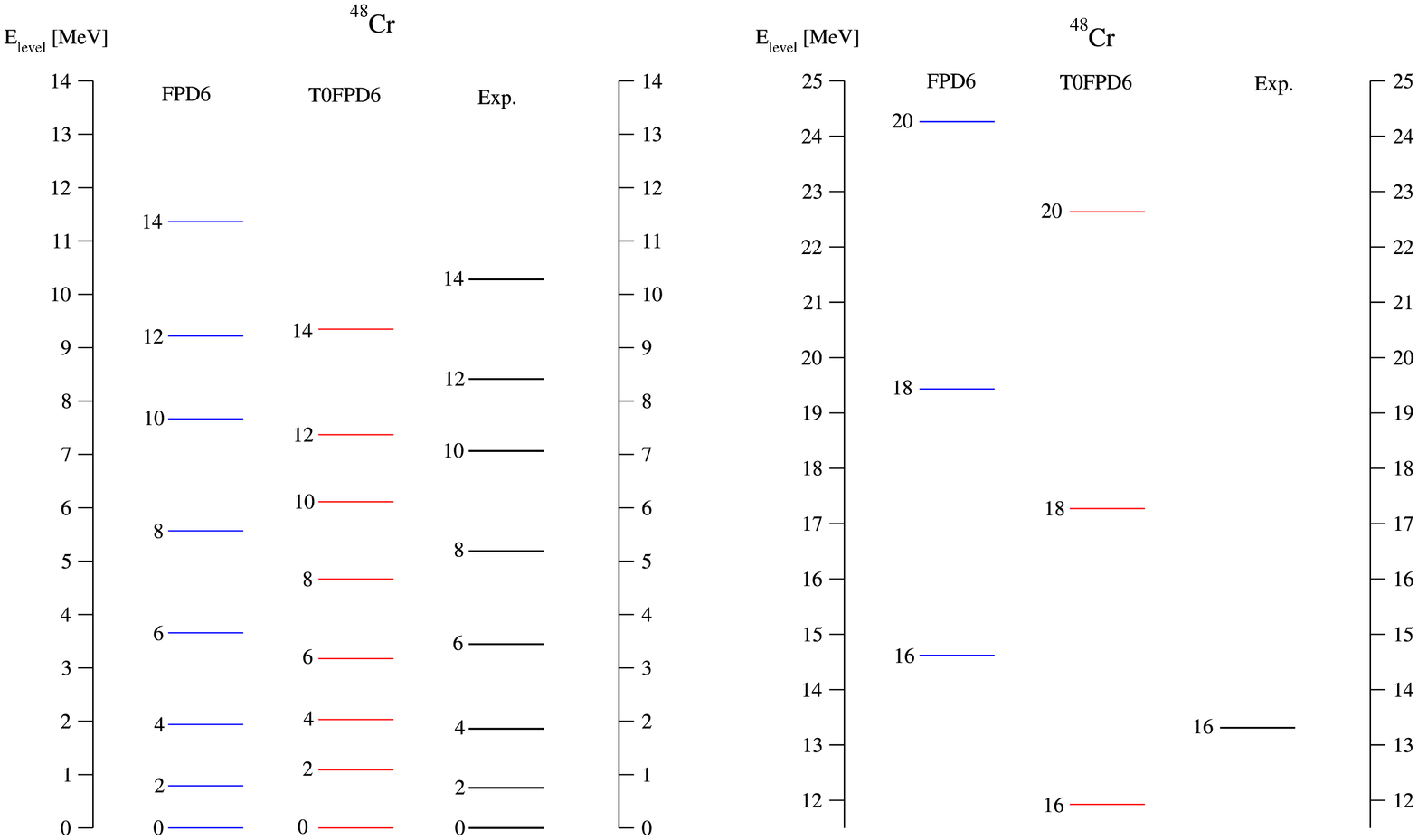}
\caption{Full fp calculations of even $J$ T=0 states in $^{48}$Cr and
comparison with experiment.
\label{fig:cr48}}
\end{figure*}

\begin{figure*}
\includegraphics[angle=270,width=170mm]{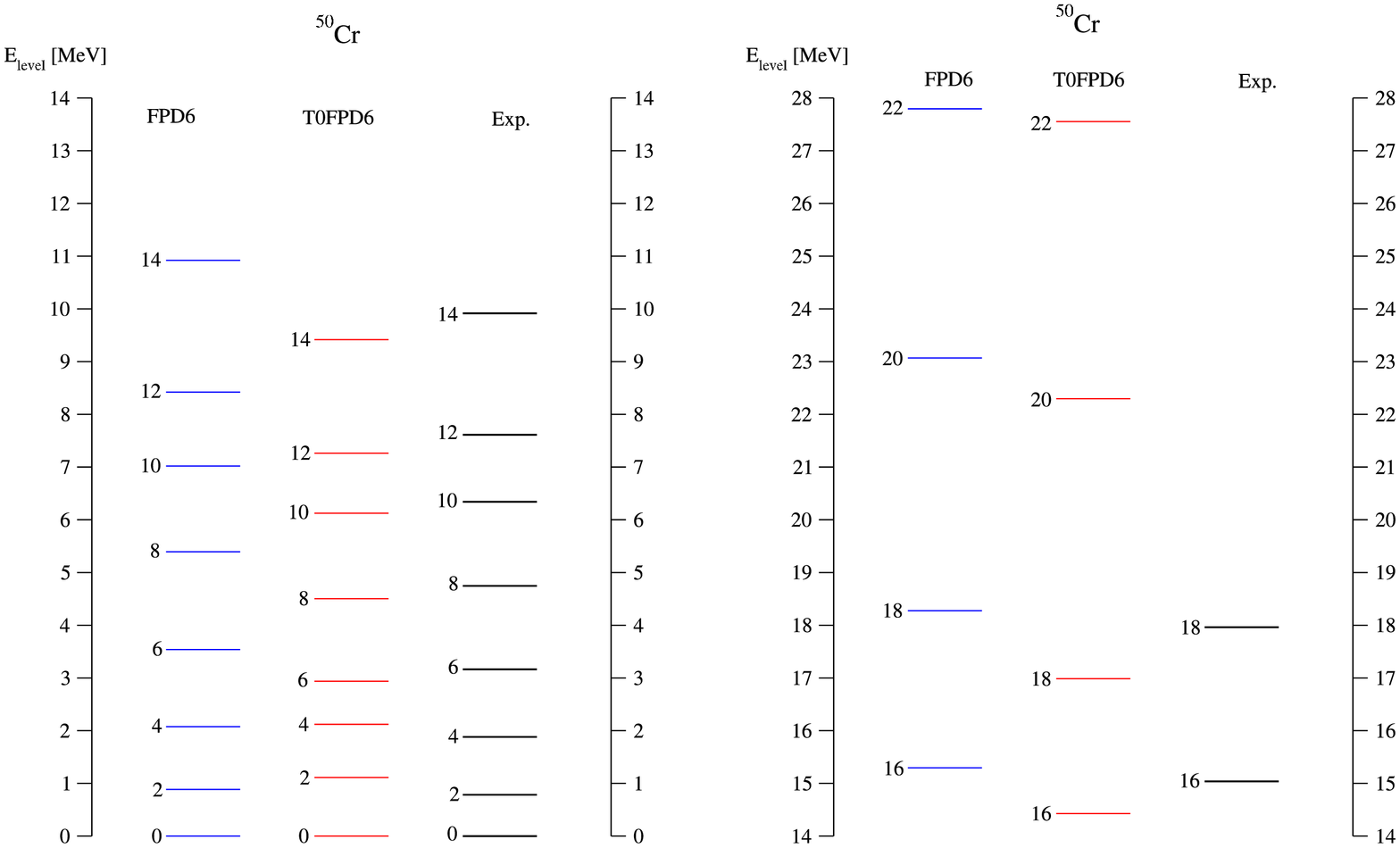}
\caption{Full fp calculations of even $J$ T=1 states in $^{50}$Cr and
comparison with experiment.
\label{fig:cr50}}
\end{figure*}

In Figures \ref{fig:ti44} to \ref{fig:cr50} we show the
T=T$_{min}$=$\frac{|N-Z|}{2}$ even $J$ states of $^{44,46,48}$Ti and
$^{48,50}$Cr. In the first column, we have the full f-p shell
calculation using FPD6.
% for $^{44,46}$Ti while for $^{48}$Ti we allow up to 4 nucleons to be excited 
% from the f$_{7/2}$ to the rest of the f-p shell (t=4). 
In the second column, we have T0FPD6, which signifies
that the T=0 two-body matrix elements have been set to zero. In the
third column, experimental yrast levels are shown~\cite{web}. 
We show separately a comparison of
the odd $J$ states in the Ti and Cr isotopes for FPD6 and T0FPD6 in Figures
\ref{fig:ti44odd} to \ref{fig:cr50odd}.
These figures show experimental even $J$ levels and some odd $J$ levels, 
although not much is known about the odd
$J$ states in these nuclei. Hopefully this paper will serve as an
impetus to search for such states. 

We will now make some broad remarks about the results. The first point to be 
made is that with the full FPD6 interaction one
gets a very good overall fit to the experimental spectrum. This should
not come as a surprise. They were designed to do so.

What is surprising is that, when we set all T=0 matrix elements to zero 
(T0FPD6), we get a semi-resonable spectrum. If this were not the case, then
what we are doing would make no sense. It would appear that the T=1 two-body 
matrix elements, acting alone, gave us the ``spine'' of the spectrum. The
addition of the T=0 matrix elements gives a needed overall improvement, but
as we will show later, there are still some discrepancies even with the full
interaction.

I should be pointed out that if we had reversed the procedure and set all the
T=1 matrix elements to zero and kept the T=0 matrix elements as they are, we
would get an unrecognizably bad spectrum.

%What is surprising is that when we set all T=0 matrix elements to zero
%as in T0FPD6 we do not get chaotic results---the spectra look
%reasonable both in comparison with the full FPD6 and experiment. Indeed
%some individual levels are closer to experiment with T0FPD6 than with
%FPD6. As expected, the fit is somewhat better with FPD6. We are however
%justified in saying that the T=1 part of the FPD6 interaction, acting
%alone, gives us the ``spine'' of the spectrum which the addition of the
%T=0 part then fine tunes. As mentioned before, a T=0 monopole
%interaction will not affect the spectrum.

We next take a closer look at the two calculated spectra. We see that
with the full FPD6 in $^{44}$Ti, the spectra for $J=0, 2, 4, 6$, and 8
looks somewhat more rotational than with T0FPD6. This is consistent
with the knowledge that the T=0 n-p interaction enhances the nuclear
collectivity. In the rotational limit the spectrum would be of the
form $J(J+1)$ while in the simple vibrational limit one gets equally
spaced levels. Experiment resides between these two limits.

Comparing FPD6 to T0FPD6, we find a closer agreement for the even J
spectra for $^{46}$Ti than for $^{44}$Ti. Indeed the closeness in
$^{46}$Ti is remarkable. It could be that $^{44}$Ti is relatively more
rotational than $^{46}$Ti and hence the T=0 interaction plays a more
important role. This could also be a measure of the relative numbers
of T=0 pairs in a T=0 nucleus as opposed to a T=1 nucleus.

With one notable exception, the spectrum of $^{48}$Ti is as good as
that of $^{46}$Ti when the T=0 matrix elements are set to zero. The
exception concerns the ($J=6$, $J=4$) splitting which is too small when we
remove the T=0 matrix elements. This could be connected with the near
degeneracy of the two lowest $6^+$ levels in $^{48}$Ti, a problem that we
previously addressed in~\cite{486dow}. In FPD6 these levels are
separated by 0.08 MeV and in T0FPD6 by 0.23 MeV. In the single j shell
model, the two $J=6^+$ states have opposite signatures---this might explain in
part why there is not a lot of level repulsion between both $J=6^+$
states.

For the even $J$ states of the N=Z nucleus $^{48}$Cr, the low spin spectrum
($J=0,2,4$, and 6) is more in the direction of a rotational spectrum
with FPD6 than it is with T0FPD6. At higher spins the FPD6 states are
at a higher energy than those of T0FPD6. For example, there is a
substantial difference---almost 2~MeV for the $J=14^{+}$
state. However it should be noted that the \textbf{experimental} value of
the energy of the $J=14^{+}$ state is in between that of the T0FPD6
and FPD6 calculations and is actually closer to T0FPD6. Similiar
results hold for $J=8, 10, 12$, and 16.

For $^{50}$Cr there is a similiar story but the differences are not so
pronounced. It is difficult to say for which of the two interactions, FPD6
or T0FPD6, the agreement with the experimental spectrum is better.

As an overview, if we look at the results for all the even--even nuclei, we 
find that the full FPD6 interaction somewhat goes too far in the description
of rotational motion, but T0FPD6 does not go far enough. This is especially
evident by looking at the high spin states, which, on the average, are too
high with FPD6 but too low with T0FPD6. The experimental energies are between
these two limits.

\subsection{Odd $J$ states in even--even Ti and Cr isotopes}

We show a comparison between FPD6 and T0FPD6 in Figures
\ref{fig:ti44odd} to \ref{fig:cr50odd} for the odd $J^+$ excitation
energies in $^{44}$Ti,$^{46}$Ti, $^{48}$Ti, $^{48}$Cr and
$^{50}$Cr. We note that the experimental data on odd $J$ is very sparse.
%We do not compare with experiment as the data on odd J is very sparse.  
In $^{44}$Ti there are no odd $J$, T=0 states
identified. There is a known 1$^+$ state at 7216 keV but this state
has isospin 1 and has been associated with the scissors mode state.
In $^{46}$Ti there are 2 nearly degenerate 1$^+$ T=1 states at 3731
and 3872 MeV. In the f-p shell model space one can only get one 1$^+$
state at this energy; one of these must be an intruder state. In general,
as stated above, there are not too many odd $J$, T=$|N-Z|/2$ states known in 
the even--even Ti and Cr isotopes. We show in the relevant figures the few that
are known.

%There is an 11$^+$ state at 7941.8 keV.  In $^{48}$Ti the lowest 3$^+$,
%1$^+$, and 7$^+$ states are known to be at 3224, 3738, and 5196 keV
%respectively.

In $^{44}$Ti the ordering of odd $J$ states is the same for T0FPD6 as it
is for FPD6. However there is a large overall downward shift. This can
be taken care of by a one-body field. When this is done the comparison
is fairly good but there are some deviations. The splitting of the
$J=1^+$ and $11^+$ states (neither of these sates is present in the
f$_{7/2}$ model space) is much larger for T0FPD6 than for FPD6. There
is more sensitivity in the odd $J$ spectrum to the T=0 two-body matrix
elements than for even $J$.  It would therefore be worthwhile to devise
experiments that can find these odd $J$ states. 
% presumably one needs projectiles with spin.

In $^{46}$Ti and $^{48}$Ti the deviations between T0FPD6 and FPD6 are
not as large as for $^{44}$Ti, but there are overall one-body shifts
to be taken into account.

It should be noted that there is a simplicity in the spectrum of the
odd $J$ states . In all three Ti isotopes, we find that, except for the
$J=1^+$ state, there is a sequential ordering $J=3^+, 5^+, 7^+,
9^+, 11^+, 13^+$, and 15$^+$, which suggests a band structure
that should be investigated.

For the odd $J$ states of $^{48}$Cr, there is a downward shift in the
energies of the states calculated with T0FPD6. For J=7, the difference
is about 2 MeV. Also, very strangely, the T0FPD6 interaction, for which
the T=0 matrix elements are set to zero, gives a better fit to the
known (and admittedly incomplete) odd $J$ spin spectrum.

For the odd $J$ states of $^{50}$Cr, there is also a downward shift of
the energies when T0FPD6 is used as compared to the full FPD6
interaction. With the full interaction, there is better agreement for
the $J=1^{+}$ state, but not so for the other known states $J=5,11, 13,
15$, and 17.

One purpose of this paper is to point out that the data on odd $J$, T=T$_{min}$
states in even--even nuclei is very sparce and it would be of interest to 
device means, perhaps with radioactive beams and projectiles which have 
non-zero spin, of exciting such states and unfolding their systematics.

\subsection{The A=43 spectrum}

\begin{figure}
\includegraphics[width=85mm]{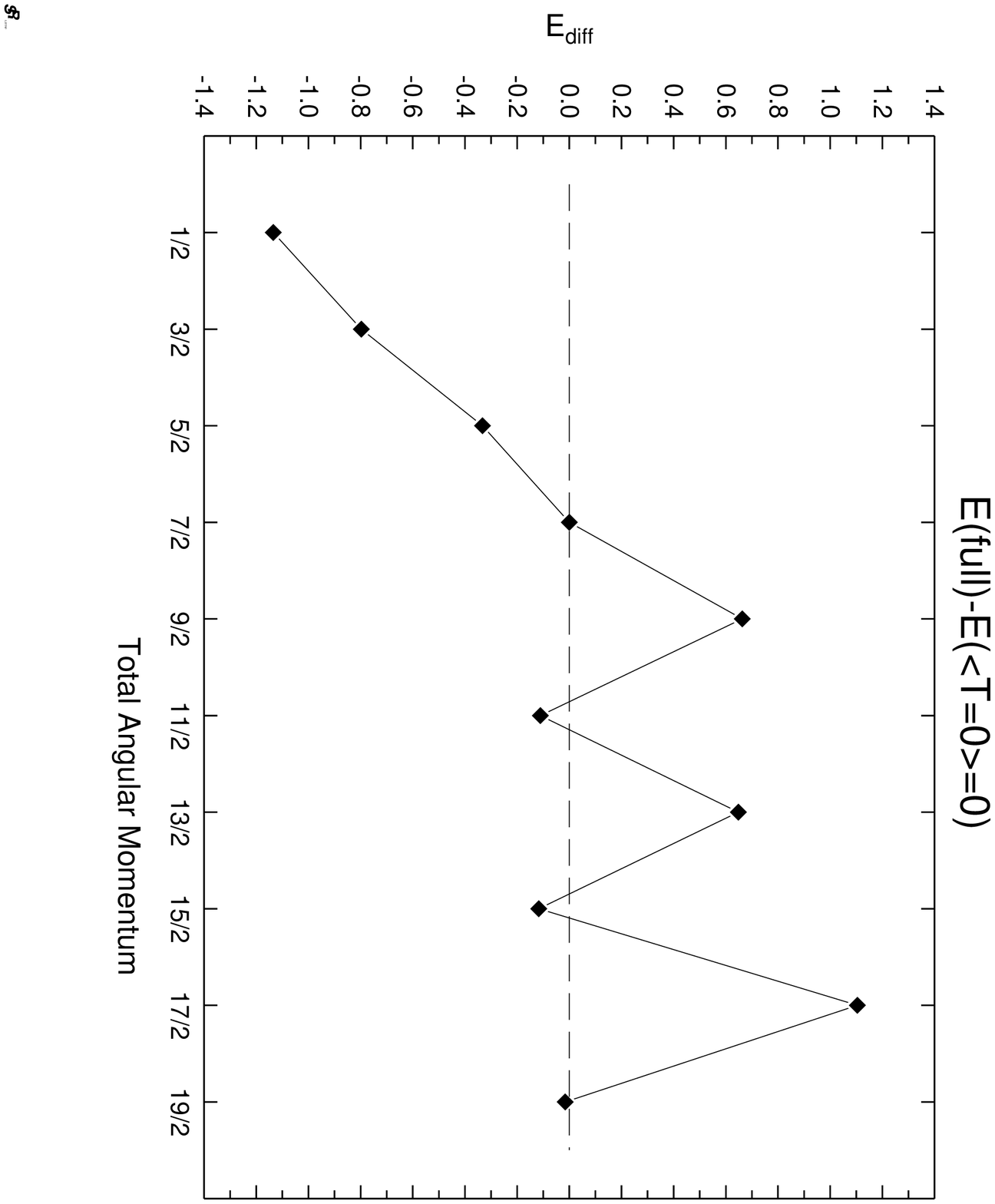}
\caption{E(FPD6)$-$E(T0FPD6) (MeV) vs Total Angular Momentum ($\hbar$)
in $^{43}$Ti.
\label{fig:ti43line}}
\end{figure}

As an example of an odd A system, consider the case of
$^{43}$Sc($^{43}$Ti) which has been previously discussed~\cite{dow1}.
We here show the figure~\ref{fig:ti43line}, which shows the difference of
the spectra when FPD6 and T0FPD6 are used. For high spins there is a
staggering with FPD6 which is not present with T0FPD6. The
$J=\frac{9}{2}$, $\frac{13}{2}$, and $\frac{17}{2}$ states come higher
than the corresponding $J=\frac{11}{2}$, $\frac{15}{2}$,
$\frac{19}{2}$ states. This staggering is due to the T=0
interaction. So, by examining this phenomenon, we could learn something
about the effective T=0 interaction in a nucleus.

\subsection{The T=0 and T=1 spectra of $^{46}$V}

Recent studies and calculations for $^{46}$V have been performed by
Mollar et~al.~\cite{mollar1} and Brandolini et~al.~\cite{brand2}.

\begin{figure}
\includegraphics[width=85mm]{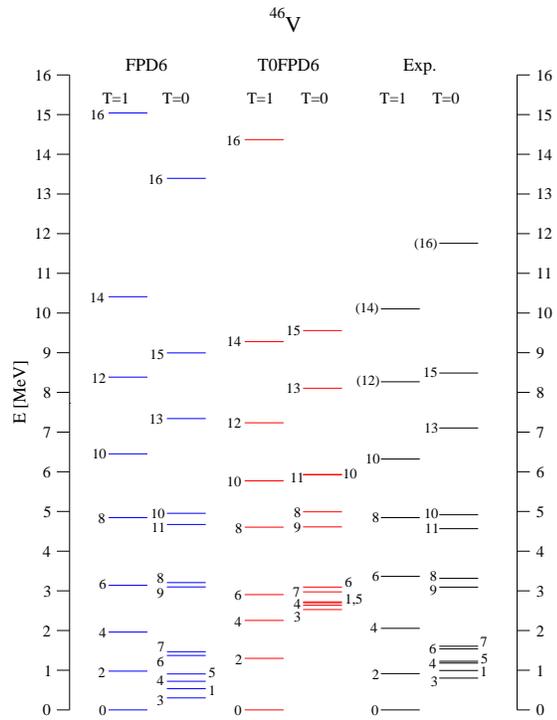}
\caption{Full fp calculation and experimental results for T=0 and 1
states in $^{46}$V.
\label{fig:46v}}
\end{figure}

In Figure \ref{fig:46v} we show a full fp calculation for the odd--odd N=Z
nucleus $^{46}$V. We show both the T=0 and T=1 states. The latter are
isobaric analog states of corresponding states in $^{46}$Ti, so the
spectra, where a charge independent interaction is used, as in this
case, are identical; thus, the previous discussion here applies.

The full FPD6 fit to experiment for the T=0 states is reasonable for
the $J=3^{+}, 4^{+}$, and 5$^{+}$ but the $J=7^{+}$ state is much
higher experimentally---it is above the 9$^{+}$ state while with FPD6
it is below. We now compare FPD6 with T0FPD6. Clearly the T=0 states
as a whole are shifted up. This can be resolved by adding the T=0
monopole interaction $a[\frac{1}{4}-t(1) \cdot t(2)]$. A downward
shift of about 1.5~MeV will make the comparison with FPD6 and
experiment much better. It is surprising that, when this shift is made,
the T=0 states agree well with FPD6 and T0FPD6.

\section{B(E2) rates}

The calculated B(E2) rates in the full fp space for $^{44}$Ti, $^{46}$Ti, 
$^{48}$Ti, $^{48}$Cr, and $^{50}$Cr are listed in Tables~\ref{tab:first} to 
\ref{tab:50cr}. 
%We allow up to $t$ nucleons to
%be excited from the f$_{7/2}$ shell to the rest of the fp shell. The
%values of $t$ used are 4, 6, and 8, respectively. 
The effective charges
used are the standard $1.5e$ for the proton and $0.5 e$ for the
neutron. The difference in the effective charges from 1 and 0 is
intended to take care of the fact that the $\Delta N =2$ and higher
excitations are not present in this model space.  The results for FPD6
and T0FPD6 are shown. We also display the ratios of the results for
the two interactions.

For $^{46}$Ti the reintroduction of the T=0 two-body matrix elements
causes an increase (relative to T0FPD6) of a factor of two or more
for all the transitions considered. So there is evidence here that the
T=0 matrix elements contribute to the collectivity. 
%This is not seen by looking at just the excitation energies in $^{46}$Ti 
%where FPD6 and T0FPD6 give very similar results.

The behavior of $^{48}$Ti is very similar to that of $^{46}$Ti with
two exceptions. The B(E2) for the transition $4 \rightarrow 6$ is clearly 
peculiar in its
behavior, as are the transitions involving the $J=12$ yrast state. While
the reason for this behavior of the $J=12$ state is not yet clear, the
$J=6$ states of $^{48}$Ti have been studied before. The existence of two
close lying 6$^{+}$ states require us to examine this closer. In Table
\ref{tab:48ti468} we examine the yrast transitions for these close
lying states, finding that it is only for the $4 \rightarrow 6_1$
transition that we get a strong enhancement when removing the T=0
matrix elements.

In the $^{48}$Cr and $^{50}$Cr (Tables~\ref{tab:48cr} and~\ref{tab:50cr}), for
the most part, the B(E2)'s are larger when the T=0 two-body matrix elements
are reintroduced, but there are some notable exceptions. In $^{48}$Cr the 
$14^+ \rightarrow 16^+$ transition is larger for T0FPD6 than for FPD6, the
ratio being 1.029. In $^{50}$Cr the ratios for $8^+ \rightarrow 10^+$,
$10^+ \rightarrow 12^+$ and $12^+ \rightarrow 14^+$ are, respectively, 2.174,
1.271, 1.116. It was previously noted by Zheng and Zamick~\cite{zz96} that the 
$10^+$ state in $^{50}$Cr is not consistent with being a member of the $K=0$ 
ground state band, rather it looked like a $K=10$ state, as noted by Zamick, 
Zheng and Fayache~\cite{zfz96}. This is in agreement with the experimental
results of Brandolini et al.~\cite{betal02}.

\section{Closing Remarks}

To partially explain why one gets a semi-reasonable spectrum with T0FPD6, we
can look at the spectrum of $^{42}$Sc, which consists of one proton and one 
neutron beyond the closed shell $^{40}$Ca. The energy levels have been used
to get a single $j$-shell two-body effective interaction in the f$_{7/2}$
shell, i.e., taking matrix elements from experiment. In this simplified
procedure, one makes the association $\langle (j^2)^J V (j^2)^J \rangle =
E(J)+\mbox{constant}$. Note that the constant will not affect the excitation
energies or wave functions in this model. Thus, for example, the excitation
energy of the $J=6^+_1$, T=1 state relative to the $J=0^+$, T=1 state is
3.122~MeV. So we have $\langle (j^2)^6 V (j^2)^6 \rangle = 3.122 \mbox{ MeV}
+\mbox{constant}$, etc.

Setting the $J=0$, T=1 energy to zero in $^{42}$Sc, the remaining states have
the following excitation energies (in MeV):

\begin{center}
\begin{tabular}{llcll}
\multicolumn{2}{c}{T=1} & \hspace{1cm} & \multicolumn{2}{c}{T=0} \\
$J$ \hspace{0.5cm} & Energy & & $J$ \hspace{0.5cm} & Energy \\
\cline{1-2} \cline{4-5}
2 & 1.613 & & 1 & 0.611 \\
4 & 2.815 & & 3 & 1.490 \\
6 & 3.122 & & 5 & 1.510 \\
 & & & 7 & 0.616 
\end{tabular}
\end{center}

Note that the total spread of the T=1 states ($(E(6)-E(0)$) is 3.122~MeV. More
than three times the spread of the T=0 states ($E(5)-E(1)$) of 0.899~MeV. 
Thus, we can say that, to a first approximation, the T=0 spectrum is almost
degenerate, judging by the scale set by the T=1 interaction. This would then
justify the starting point of setting the T=0 matrix elements to a constant.
It is easy to show that in this single $j$-shell model space, if one adds a
constant to the T=0 matrix elements, it will not affect the wave functions of
the states and will not affect the excitation energies of the states which
have the same isospin.

It can be seen that the two particle T=1 spectrum in $^{42}$Sc is quite 
different 
from that of a pairing interaction, for which the $J=2,4$ and 6 states are 
degenerate. The fact that the excitation energy of the $6^+$ state is about
twice that of the $2^+$ state indicates that other components of the
nucleon--nucleon interaction are present, e.g., a quadrupole--quadrupole
interaction. Hence, the T=1 spectrum of $^{42}$Sc has built into it some 
aspects necessary for nuclear collectivity.

The above discussion suggests that, in a full f-p calculation, the single $j$
components are sufficiently prevalent so as to get the overall pattern of the 
spectrum in reasonably good shape. The higher shell admixtures then readjust 
the spectrum so as to change from what is roughly a vibrational pattern to a
rotational one, and here the T=0 two-body matrix elements play an important
role.

In summary, in studying the problem of the T=0 neutron--proton
interaction in a nucleus, it may prove more fruitful to begin by
removing this channel altogether as was done here by setting all the
T=0 two-body matrix elements to zero and then reintroducing
them, rather than adopting the more common approach of investigating
the effects of a pairing interaction separated from the rest of the
interaction.  This may be especially true in the shell model as the
suggestion has been made by Satula and Wyss that it may not be
appropriate to separate out a pairing interaction from the rest of the
Hamiltonian in a shell model context~\cite{satw676}.

\begin{figure}
\includegraphics[width=85mm]{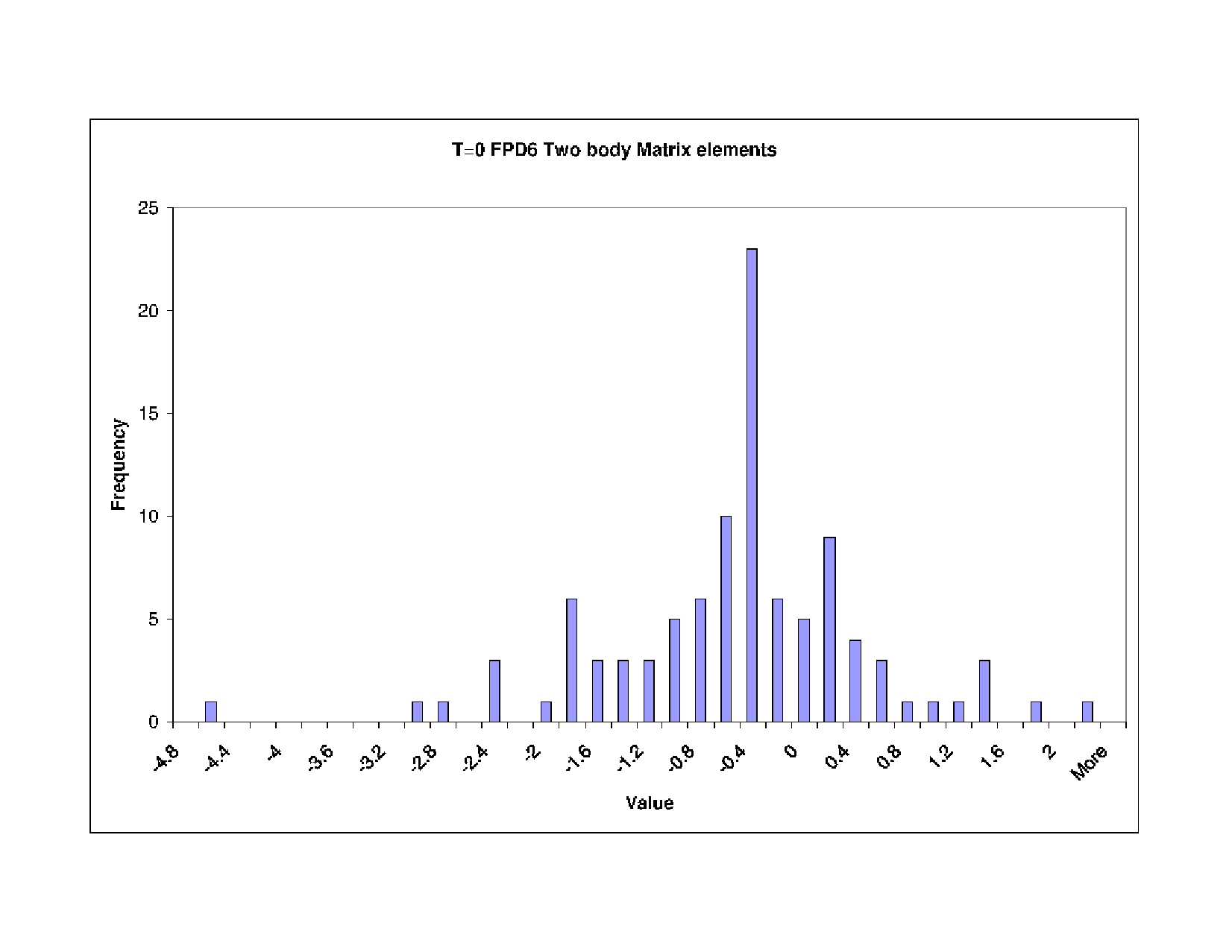}
\caption{T=0 two-body matrix element distribution for FPD6
\label{fig:newlast}}
\end{figure}

\begin{figure}
\includegraphics[width=85mm]{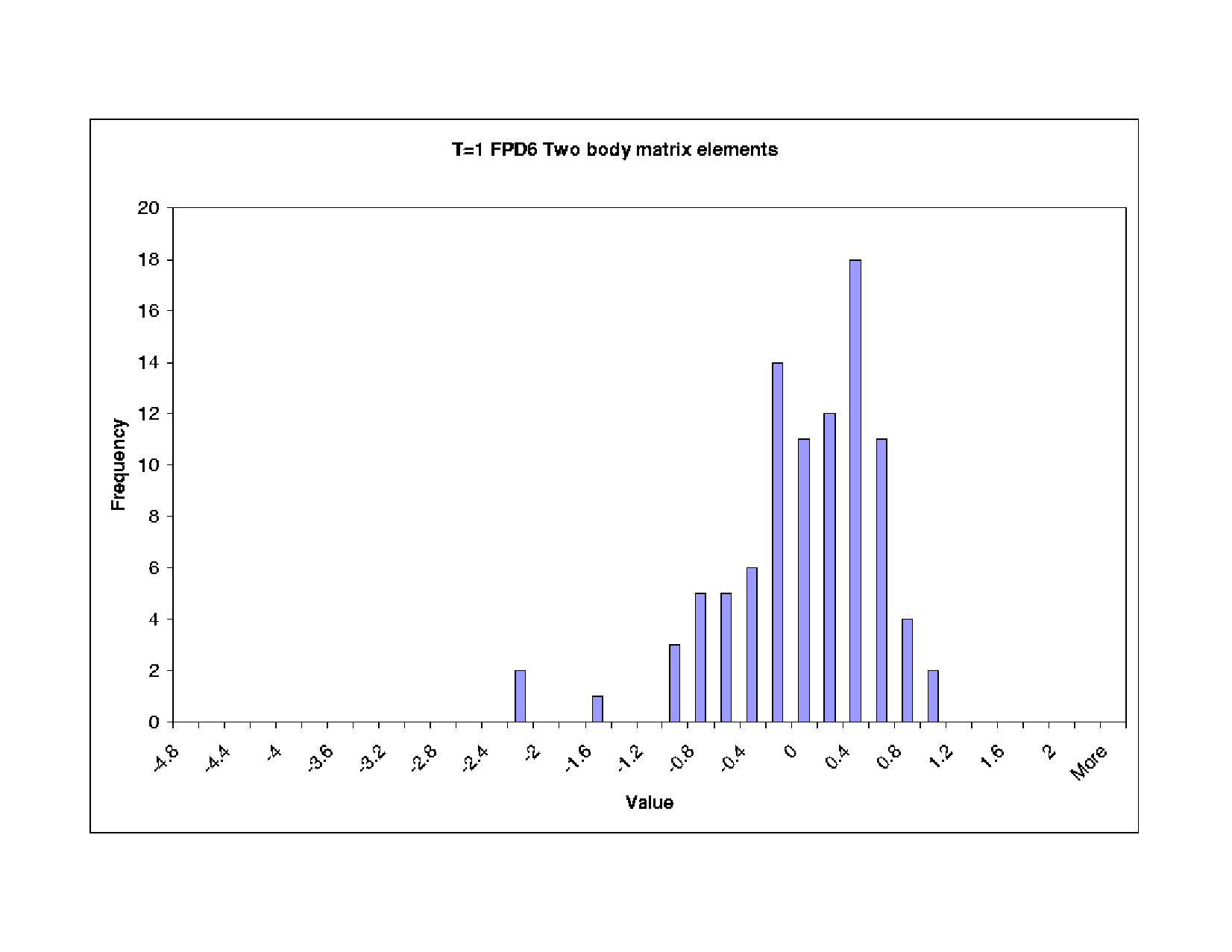}
\caption{T=1 two-body matrix element distribution for FPD6
\label{fig:finallast}}
\end{figure}

An examination of the T=0 two-body matrix elements in Figure
\ref{fig:newlast} does not show any obvious simplicity. Their
distribution looks just as complex as those with T=1 shown in Figure
\ref{fig:finallast}. If the T=0 diagonal matrix elements were all
constant and the off-diagonal matrix elements were zero, we could
represent the results by a two-body monopole interaction as $a(1/4 - t(1)
\cdot t(2))$. This would be an easy explanation of the insensitivity
but certainly it would not be a correct one.

\begin{figure}
\includegraphics[width=85mm,height=4in]{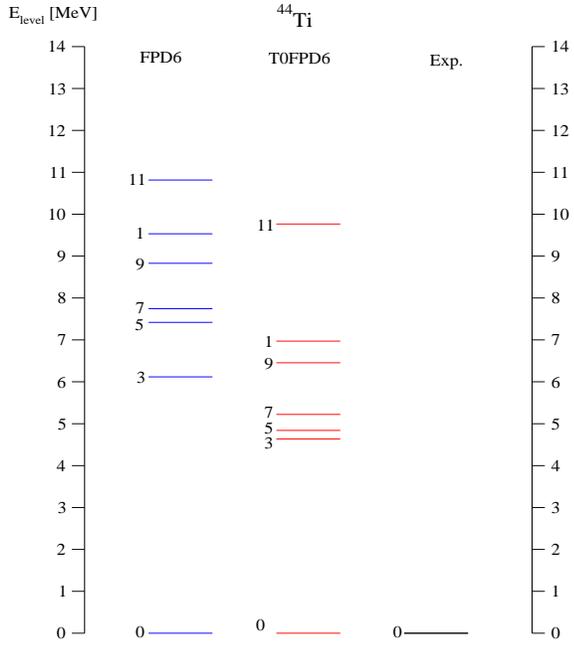}
\caption{Full fp space calculations of odd $J$ T=0 states in $^{44}$Ti.
\label{fig:ti44odd}}
\end{figure}

\begin{figure}
\includegraphics[width=85mm,height=4in]{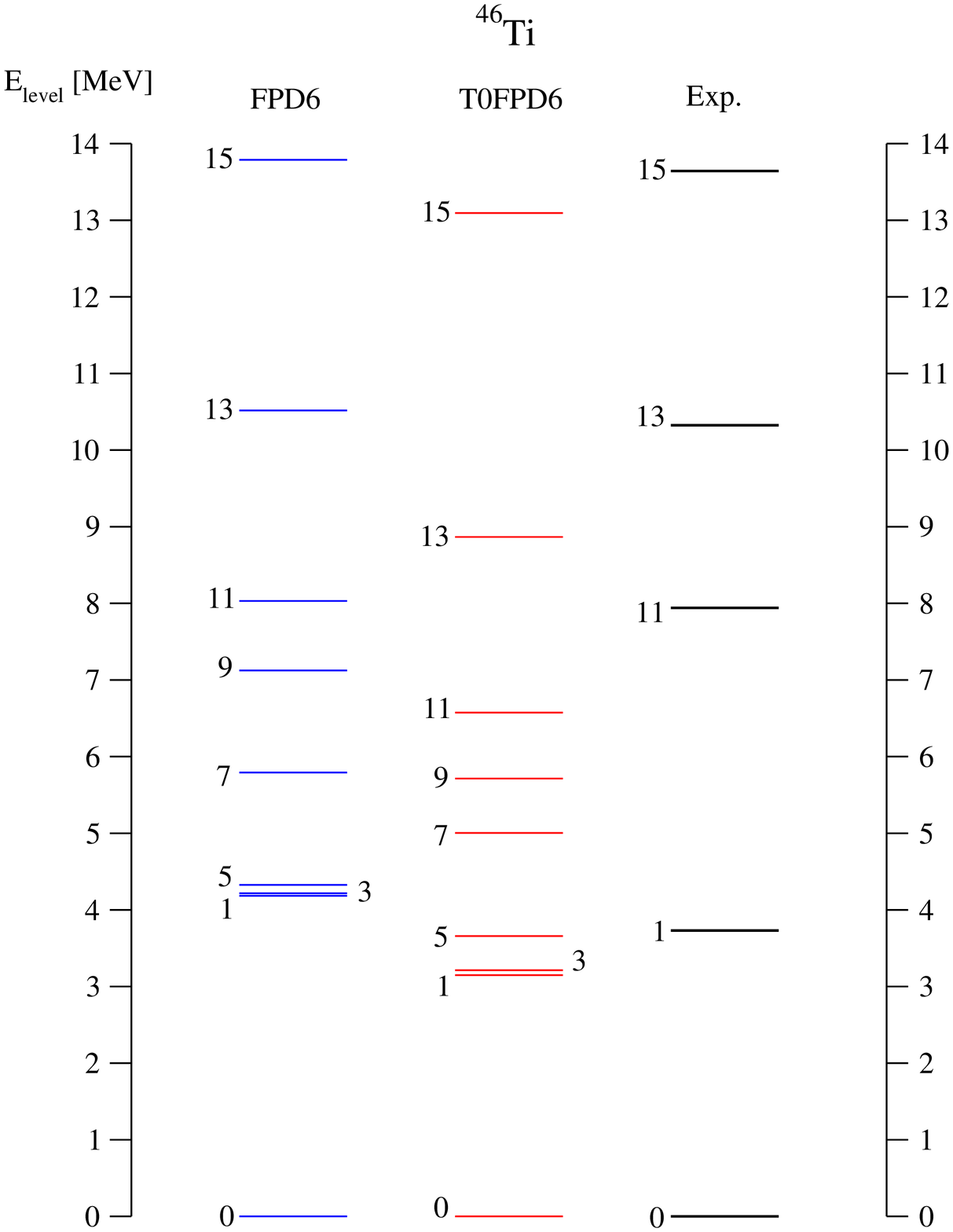}
\caption{ Full fp space calculations of odd $J$ T=1 states in $^{46}$Ti and
comparison with experiment.
\label{fig:ti46odd}}
\end{figure}

\begin{figure*}
\includegraphics[angle=270,width=170mm]{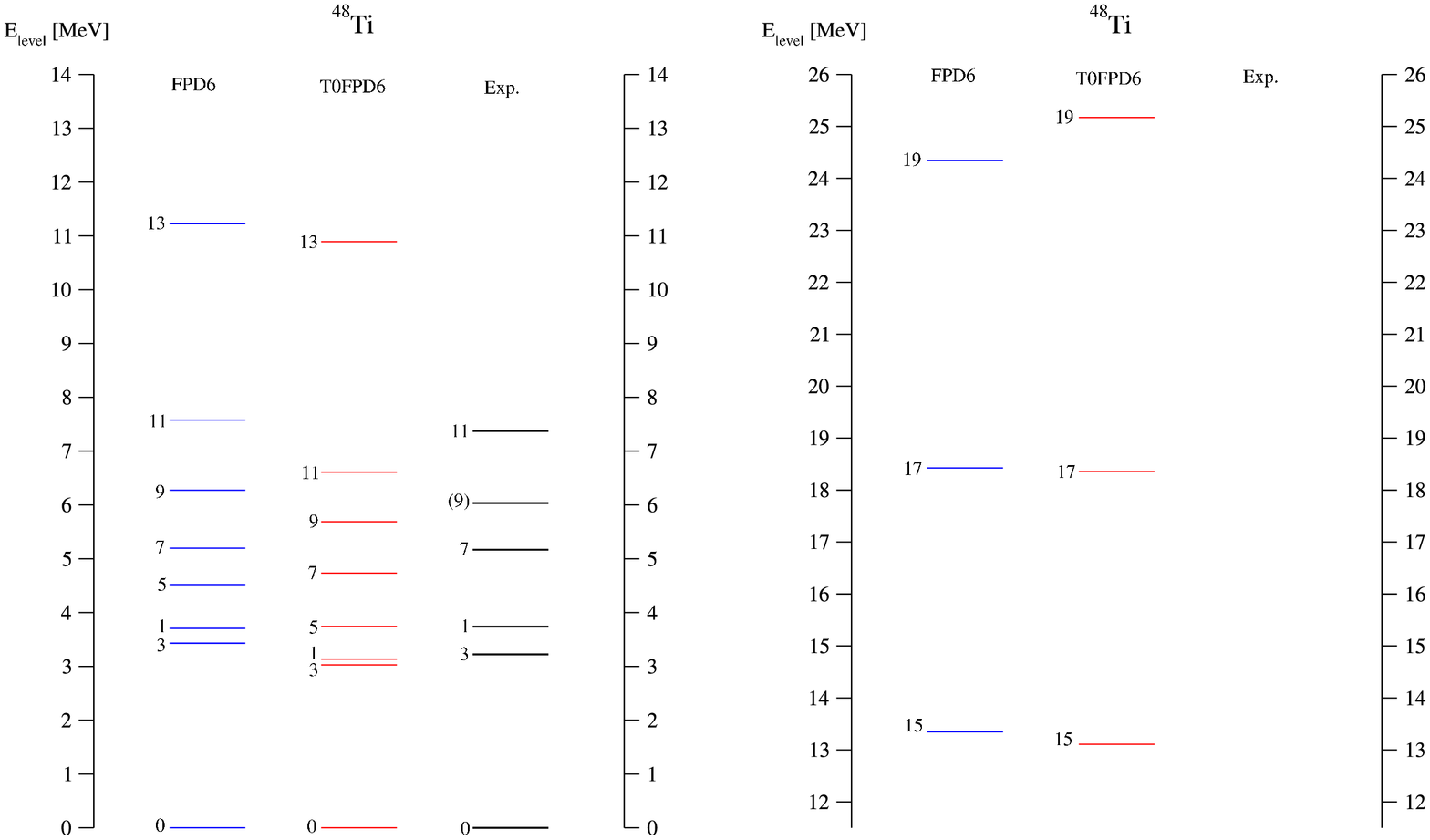}
\caption{Full fp space calculations of odd $J$ T=2 states in $^{48}$Ti
and comparison with experiment.
\label{fig:ti48odd}}
\end{figure*}

\begin{figure*}
\includegraphics[angle=270,width=170mm]{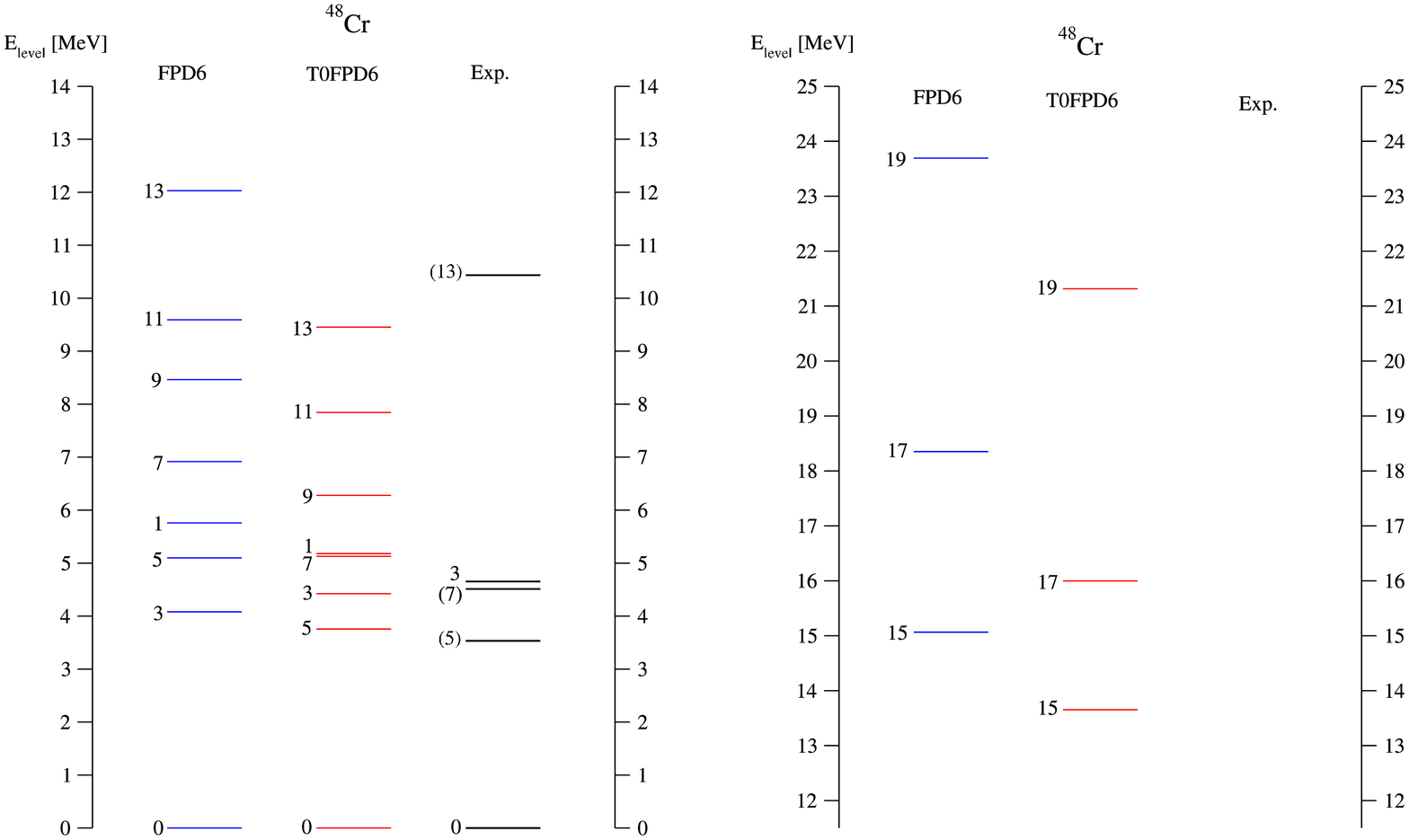}
\caption{Full fp space calculations of odd $J$ T=0 states in $^{48}$Cr
and comparison with experiment.
\label{fig:cr48odd}}
\end{figure*}

\begin{figure*}
\includegraphics[angle=270,width=170mm]{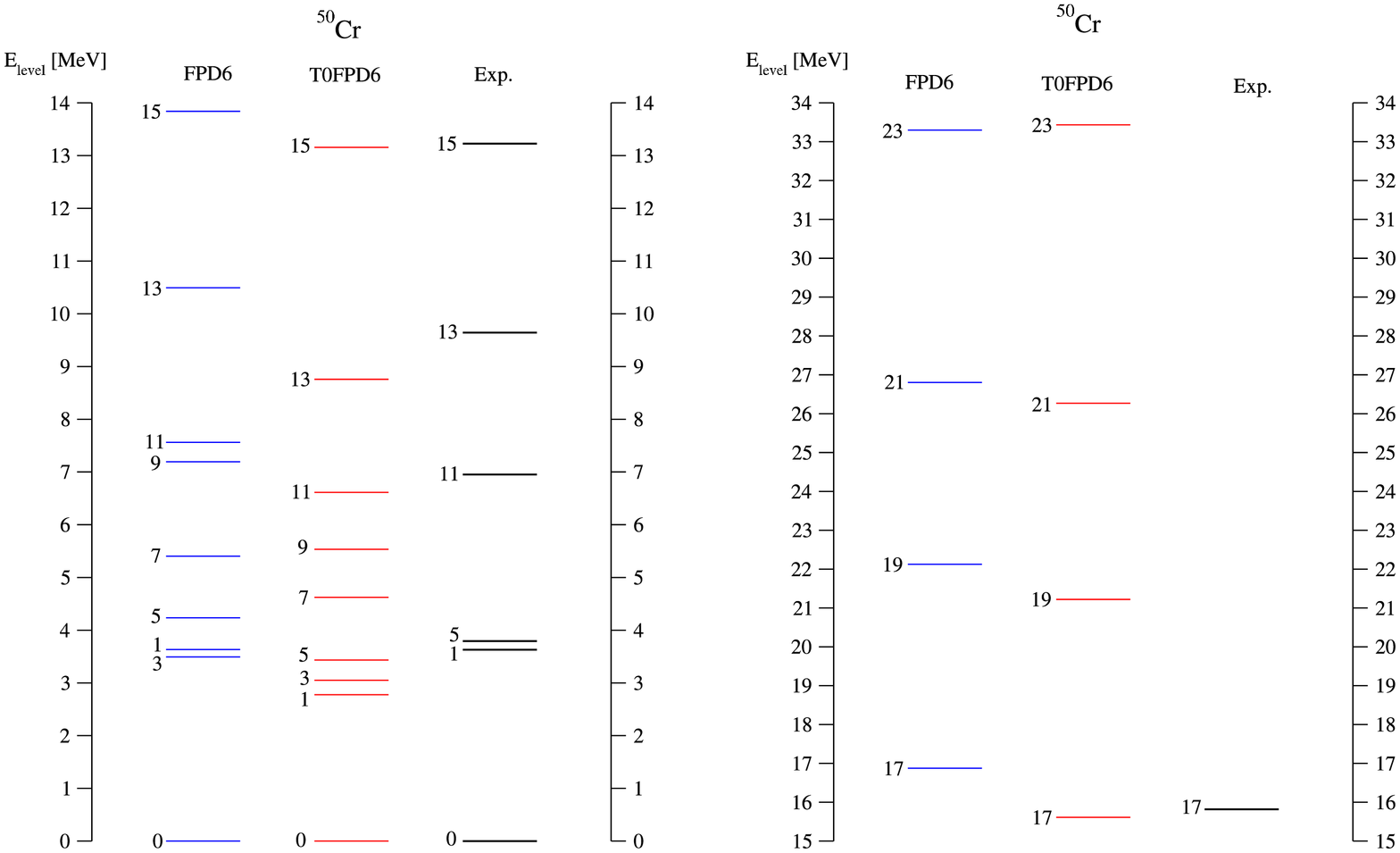}
\caption{Full fp space calculations of odd $J$ T=1 states in $^{50}$Cr
and comparison with experiment.
\label{fig:cr50odd}}
\end{figure*}

Concerning the future of this subject, it would be of great interest
to fill in the missing levels which have been shown in the tables. In
particular we have noted that, although there is much data on even
spins in the even--even nuclei, there is very little known about the odd
$J$ positive parity states. Figures \ref{fig:ti44odd} to
\ref{fig:cr50odd} show some interesting band structure for odd $J$ states.
%; in general, the full
%FPD6 interaction leads to more rotational spectra than does T0FPD6. The
%differences between FPD6 and T0FPD6 are somewhat larger for the odd $J$
%than for the even $J$ states. Hence, 
If the levels are found, we can put
more constraints on the effective nucleon--nucleon interaction in this
region.

We thank Sylvia Lenzi for her overall support and in particular for
providing us with new data on high spin states.
This work was supported by the U.S. Dept. of Energy under Grant
No. DOE FG01 04ER04-02. One of us (SJQR) would like to acknowledge
travel support from the University of Southern Indiana. A.E. is supported by 
a grant financed by the Secretar\'{\i}a de Estado de Edu\-caci\'on y 
Universidades (Spain) and cofinanced by the European Social Fund.

%\begin{figure}
%\includegraphics[width=85mm,height=4in]{ti48-spec11-even.ps}
%\caption{Full fp calculations of even $J\leq 12$ T=2 states in $^{48}$Ti and
%comparison with experiment.
%\label{fig:ti481}}
%\end{figure}

%\begin{figure}
%\includegraphics[width=85mm,height=4in]{ti48-spec12-even.ps}
%\caption{Full fp calculations of even $J>12$ T=2 states in $^{48}$Ti and
%comparison with experiment.
%\label{fig:ti482}}
%\end{figure}

%\begin{figure}
%\includegraphics[width=85mm,height=4in]{cr48-spec11-even.ps}
%\caption{Full fp calculations of even $J\leq 14$ T=2 states in ${48}$Cr and
%comparison with experiment.
%\label{fig:cr481}}
%\end{figure}

%\begin{figure}
%\includegraphics[width=85mm,height=4in]{cr48-spec12-even.ps}
%\caption{Full fp calculations of even $J>14$ T=2 states in $^{48}$Cr and
%comparison with experiment.
%\label{fig:cr482}}
%\end{figure}

%\begin{figure}
%\includegraphics[width=85mm,height=4in]{cr50-spec11-even.ps}
%\caption{Full fp calculations of even $J\leq 14$ T=2 states in $^{50}$Cr and
%comparison with experiment.
%\label{fig:cr501}}
%\end{figure}

%\begin{figure}
%\includegraphics[width=85mm,height=4in]{cr50-spec12-even.ps}
%\caption{Full fp calculations of even $J>14$ T=2 states in $^{50}$Cr and
%comparison with experiment.
%\label{fig:cr502}}
%\end{figure}

%\begin{figure}
%\includegraphics[width=85mm,height=4in]{ti48-spec11-odd.ps}
%\caption{Full fp space calculations of odd $J\leq 13$ T=2 states in $^{48}$Ti
%and comparison with experiment.
%\label{fig:ti48odd1}}
%\end{figure}

%\begin{figure}
%\includegraphics[width=85mm,height=4in]{ti48-spec12-odd.ps}
%\caption{Full fp space calculations of odd $J\geq 15$ T=2 states in $^{48}$Ti.
%\label{fig:ti48odd2}}
%\end{figure}

%\begin{figure}
%\includegraphics[width=85mm,height=4in]{cr48-spec11-odd.ps}
%\caption{Full fp space calculations of odd $J\leq 13$ T=2 states in $^{48}$Cr
%and comparison with experiment.
%\label{fig:cr48odd1}}
%\end{figure}

%\begin{figure}
%\includegraphics[width=85mm,height=4in]{cr48-spec12-odd.ps}
%\caption{Full fp space calculations of odd $J\geq 15$ T=2 states in $^{48}$Cr.
%\label{fig:cr48odd2}}
%\end{figure}

%\begin{figure}
%\includegraphics[width=85mm,height=4in]{cr50-spec11-odd.ps}
%\caption{Full fp space calculations of odd $J\leq 13$ T=2 states in $^{50}$Cr
%and comparison with experiment.
%\label{fig:cr50odd1}}
%\end{figure}

%\begin{figure}
%\includegraphics[width=85mm]{cr50-spec12-odd.ps}
%\caption{Full fp space calculations of odd $J\geq 15$ T=2 states in $^{50}$Cr
%and comparison with experiment.
%\label{fig:cr50odd2}}
%\end{figure}

\begin{table}
\begin{center}
\caption{$^{44}$Ti yrast B(E2) values [e$^2$fm$^4$] in full FPD6 and T0FPD6}
\label{tab:first}
\begin{tabular}{ccccccc}
\hline \hline \\
Transition & & FPD6 & & T0FPD6 & & ratio \\
\\
\hline \hline \\
 & & & & \\
$0 \rightarrow 2$ & & 607.24 & & 375.09 & & 0.618 \\
$2 \rightarrow 4$ & & 297.71 & & 146.18 & & 0.491 \\
$4 \rightarrow 6$ & & 202.05 & & 61.164 & & 0.303 \\
$6 \rightarrow 8$ & & 127.20 & & 65.242 & & 0.513 \\
$8 \rightarrow 10$ & & 117.50 & & 78.088 & & 0.665 \\
$10 \rightarrow 12$ & & 65.501 & & 47.968 & & 0.732 \\
\\
\hline \hline
\end{tabular}
\end{center}
\end{table}

\begin{table}
\begin{center}
\caption{$^{46}$Ti yrast B(E2) values [e$^2$fm$^4$] in full FPD6 and T0FPD6}
\label{tab:second}
\begin{tabular}{ccccccc}
\hline \hline \\
Transition & & FPD6 & & T0FPD6 & & ratio \\
\\
\hline \hline \\
 & & & & \\
$0 \rightarrow 2$ & & 682.06 & & 432.81 & & 0.635 \\
$2 \rightarrow 4$ & & 349.03 & & 179.18 & & 0.513 \\
$4 \rightarrow 6$ & & 273.85 & & 92.867 & & 0.339 \\
$6 \rightarrow 8$ & & 218.61 & & 82.478 & & 0.377 \\
$8 \rightarrow 10$ & & 157.63 & & 75.154 & & 0.477 \\
$10 \rightarrow 12$ & & 56.441 & & 29.610 & & 0.525 \\
$12 \rightarrow 14$ & & 39.923 & & 18.930 & & 0.474 \\
$14 \rightarrow 16$ & & 1.1333 & & 0.4274 & & 0.377 \\
\\
\hline \hline
\end{tabular}
\end{center}
\end{table}

\begin{table}
\begin{center}
\caption{$^{48}$Ti yrast B(E2) values [e$^2$fm$^4$] in full FPD6 and T0FPD6}
\label{tab:new}
\begin{tabular}{ccccccc}
\hline \hline \\
Transition & & FPD6 & & T0FPD6 & & ratio \\
\\
\hline \hline \\
 & & & & \\
$0 \rightarrow 2$ & & 560.78 & & 401.97 & & 0.717 \\
$2 \rightarrow 4$ & & 306.35 & & 171.89 & & 0.561 \\
$4 \rightarrow 6$ & & 64.147 & & 76.029 & & 1.185 \\
$6 \rightarrow 8$ & & 79.337 & & 26.664 & & 0.336 \\
$8 \rightarrow 10$ & & 75.571 & & 39.341 & & 0.521 \\
$10 \rightarrow 12$ & & 30.055 & & 29.710 & & 0.988 \\
$12 \rightarrow 14$ & & 5.0445 & & 3.4293 & & 0.680 \\
$14 \rightarrow 16$ & & 42.526 & & 11.608 & & 0.273 \\
$16 \rightarrow 18$ & & 0.9308 & & 0.3383 & & 0.363 \\
\\
\hline \hline
\end{tabular}
\end{center}
\end{table}

\begin{table}
\begin{center}
\caption{$^{48}$Ti yrast B(E2) values [e$^2$fm$^4$] in full FPD6 and T0FPD6
for $6^+$ states}
\label{tab:48ti468}
\begin{tabular}{ccccccc}
\hline \hline \\
Transition & & FPD6 & & T0FPD6 & & ratio \\
\\
\hline \hline \\
 & & & & \\
$4 \rightarrow 6_1$ & & 64.147 & & 76.195 & & 1.188 \\
$4 \rightarrow 6_2$ & & 129.29 & & 7.3963 & & 0.057 \\
$6_1 \rightarrow 8$ & & 79.337 & & 26.570 & & 0.335 \\
$6_2 \rightarrow 8$ & & 37.301 & & 14.443 & & 0.387 \\
\\
\hline \hline
\end{tabular}
\end{center}
\end{table}

\begin{table}
\begin{center}
\caption{$^{48}$Cr yrast B(E2) values [e$^2$fm$^4$] in full FPD6 and T0FPD6}
\label{tab:48cr}
\begin{tabular}{ccccccc}
\hline \hline \\
Transition & & FPD6 & & T0FPD6 & & ratio \\
\\
\hline \hline \\
 & & & & \\
$0 \rightarrow 2$ & & 1378.4 & & 813.06 & & 0.590 \\
$2 \rightarrow 4$ & & 692.96 & & 376.46 & & 0.543 \\
$4 \rightarrow 6$ & & 577.42 & & 230.16 & & 0.399 \\
$6 \rightarrow 8$ & & 491.87 & & 241.58 & & 0.491 \\
$8 \rightarrow 10$ & & 371.28 & & 194.37 & & 0.523 \\
$10 \rightarrow 12$ & & 157.33 & & 123.28 & & 0.784 \\
$12 \rightarrow 14$ & & 140.42 & & 112.80 & & 0.803 \\
$14 \rightarrow 16$ & & 69.141 & & 71.157 & & 1.029 \\
$16 \rightarrow 18$ & & 1.8306 & & 1.4921 & & 0.815 \\
$18 \rightarrow 20$ & & 7.5903 & & 1.8787 & & 0.247 \\
\\
\hline \hline
\end{tabular}
\end{center}
\end{table}

\begin{table}
\begin{center}
\caption{$^{50}$Cr yrast B(E2) values [e$^2$fm$^4$] in full FPD6 and T0FPD6}
\label{tab:50cr}
\begin{tabular}{ccccccc}
\hline \hline \\
Transition & & FPD6 & & T0FPD6 & & ratio \\
\\
\hline \hline \\
 & & & & \\
$0 \rightarrow 2$ & & 1219.0 & & 736.60 & & 0.604 \\
$2 \rightarrow 4$ & & 636.22 & & 341.01 & & 0.536 \\
$4 \rightarrow 6$ & & 427.64 & & 147.30 & & 0.344 \\
$6 \rightarrow 8$ & & 349.16 & & 156.47 & & 0.448 \\
$8 \rightarrow 10$ & & 36.549 & & 79.449 & & 2.174 \\
$10 \rightarrow 12$ & & 48.488 & & 61.638 & & 1.271 \\
$12 \rightarrow 14$ & & 66.120 & & 73.792 & & 1.116 \\
$14 \rightarrow 16$ & & 4.3417 & & 3.4128 & & 0.786 \\
$16 \rightarrow 18$ & & 85.995 & & 42.408 & & 0.493 \\
$18 \rightarrow 20$ & & 1.8424 & & 0.8246 & & 0.448 \\
\\
\hline \hline
\end{tabular}
\end{center}
\end{table}

%\begin{table}
%\begin{center}
%\caption{$^{48}$Cr yrast B(E2) values ($e^{2}fm^{4}$) in FPD6
% and T0FPD6} 
%\label{tab:third}
%\begin{tabular}{cccc}
%&t=2 &t=2& ratio \\ 
%&FPD6& T0FPD6&  \\
%0 $\rightarrow$ 2  & 892   &691.6 &0.775\\
%2 $\rightarrow$ 4  & 424.7 &287.8 &0.678\\
%4 $\rightarrow$ 6  & 345   &169.4 &0.491\\
%6 $\rightarrow$ 8  & 319.5 &202.1 &0.633\\
%8 $\rightarrow$ 10 & 238.9 &175.9 &0.736\\
%10 $\rightarrow$ 12& 168.9 &133   &0.787\\
%12 $\rightarrow$ 14& 138   &120.7 &0.875\\
%14 $\rightarrow$ 16& 77.60 &79.22 &1.021\\
%16 $\rightarrow$ 18& 1.178 &1.471 &1.249\\
%18 $\rightarrow$ 20& 2.555 &0.728 &0.285\\
%\end{tabular}
%\end{center}
%\end{table}

%\begin{table}
%\begin{center}
%\caption{$^{50}$Cr yrast B(E2) values ($e^{2}fm^{4}$) in FPD6
% and T0FPD6} 
%\label{tab:last}
%\begin{tabular}{cccc}
%&t=2 &t=2 & ratio \\ 
%&FPD6& T0FPD6&  \\
%0 $\rightarrow$ 2  & 761.8 & 614.6 & 0.807\\
%2 $\rightarrow$ 4  & 396.8 & 271.9   & 0.685 \\
%4 $\rightarrow$ 6  & 207.3   & 116.9   & 0.564\\
%6 $\rightarrow$ 8  & 188.3 & 133   & 0.706\\
%8 $\rightarrow$ 10 & 46.48 & 69.12 & 1.486\\
%10 $\rightarrow$ 12& 54.83 & 63.79 & 1.163\\
%12 $\rightarrow$ 14& 74.58 & 80.91 & 1.085\\
%14 $\rightarrow$ 16& 6.665 & 3.647 & 0.547\\
%16 $\rightarrow$ 18& 86.84 & 36.68 & 0.422\\
%18 $\rightarrow$ 20& 0.8343& 1.006 & 1.206\\
%\end{tabular}
%\end{center}
%\end{table}


\begin{thebibliography}{99}

\bibitem{goodmannew} Alan L. Goodman, Phys. Rev. \textbf{C60}, 014311
(1999); \textbf{C63}, 044325 (2001).

\bibitem{macc1} A.O. Macchiavelli \textit{et al.}, Phys. Rev. \textbf{C61},
041303 (2000).

\bibitem{8488} N. Marginean \textit{et al.}, Phys. Rev.  \textbf{C65}, 051303
(2002).

\bibitem{antoine}E. CAURIER, shell model code ANTOINE, IRES,
STRASBOURG 1989-2004; E. CAURIER, F. NOWACKI Acta Physica Polonica 30
(1999) 705.

\bibitem{wrichter1} W.A. Richter M.G. Van Der Merwe, R.E. Julies, and
B.A. Brown, Nuc. Phys. \textbf{A523}, 325 (1991).

\bibitem{satula1} W. Satula, D.J. Dean, J. Gary, S. Mizutori, and
W. Nazarewicz, Phys. Lett. \textbf{B407}, 103 (1997).

\bibitem{bf64} R.K.~Bansal and J.B.~French, Phys. Lett. {\bf 11}, 145 (1964).

\bibitem{z65} L.~Zamick, Phys. Lett. {\bf 19}, 580 (1965).

\bibitem{chas} R..R. Chasman, Phys. Lett. \textbf{B553}, 204 (2003).

\bibitem{dow1} S.J.Q. Robinson and Larry Zamick, Phys. Rev. 
\textbf{C63}, 064316 (2001).

\bibitem{dow2} S.J.Q. Robinson and Larry Zamick, Phys. Rev. 
\textbf{C64}, 057302 (2001).

\bibitem{dow3}S.J.Q. Robinson, Ph.D. Thesis, Rutgers University (2002).

\bibitem{dow4}S.J.Q. Robinson and Larry Zamick, Phys. Rev. 
\textbf{C66}, 034303 (2002).

\bibitem{web} Data extracted using the NNDC website www.nndc.bnl.gov
from the ENSDF database

\bibitem{486dow}  S.J.Q. Robinson and Larry Zamick, Phys. Rev.
\textbf{C63}, 057301 (2001).

\bibitem{mollar1}A. Mollar et. al. Phys. Rev \textbf{C67}, 011301(R)
(2003).

\bibitem{brand2}F.Brandolini et. al. Phys. Rev. \textbf{C64} 044307
(2004)

\bibitem{satw676}W. Satula and R. Wyss, Nuc. Phys. \textbf{A676}, 120
(2000).

\bibitem{zz96} L.~Zamick and D.C.~Zheng, Phys. Rev. \textbf{C54}, 956 (1996).

\bibitem{zfz96} L.~Zamick, M.~Fayache and D.C.~Zheng, Phys. Rev. 
\textbf{C53}, 188 (1996).

\bibitem{betal02} F.~Brandolini et al., Phys. Rev. \textbf{C66}, 024304 (2002).

\end{thebibliography}
\end{document}